\pdfoutput=1
\documentclass[journal,twoside,web]{ieeecolor_arxiv}
\usepackage{tmi}
\usepackage{cite}
\usepackage{mathtools}
\usepackage{amsmath,amssymb,amsfonts}
\usepackage{booktabs}
\usepackage{graphicx}
\usepackage{textcomp}
\usepackage{float}
\usepackage{subcaption}
\usepackage{algorithm}
\usepackage{algpseudocode}
\usepackage{caption}

\graphicspath{{./trans_fig/}}
\usepackage{adjustbox}
\usepackage[pagebackref,breaklinks,colorlinks]{hyperref}

\def\BibTeX{{\rm B\kern-.05em{\sc i\kern-.025em b}\kern-.08em
    T\kern-.1667em\lower.7ex\hbox{E}\kern-.125emX}}
\begin{document}
\title{A Three-Player GAN for Super-Resolution in Magnetic Resonance Imaging}
\author{Qi Wang, Lucas Mahler, Julius Steiglechner, Florian Birk, Klaus Scheffler, Gabriele Lohmann
\thanks{Julius Steiglechner and Florian Birk are supported by the Deutsche Forschungsgemeinschaft(DFG – German Research Foundation). The grant number for J.S. is DFG LO1728/2-1 and the grant number for F.B. is DFG HE 9297/1-1.}
\thanks{Qi Wang, Lucas Mahler, and Julius Steiglechner are with Max Planck Institute for Biological Cybernetics, T\"ubingen, Germany(e-mail:qi.wang@tuebingen.mpg.de; lucas.mahler@tuebingen.mpg.de; julius.steiglechner@tuebingen.mpg.de).}
\thanks{Florian Birk, Klaus Scheffler and Gabriele Lohmann are  with both Max Planck Institute for Biological Cybernetics and University Hospital of T\"ubingen, T\"ubingen, Germany(e-mail:florian.birk@tuebingen.mpg.de; klaus.scheffler@tuebingen.mpg.de; lohmann@tuebingen.mpg.de)}}
\maketitle

\begin{abstract}
    Learning based single image super resolution (SISR) task is well investigated in 2D images. However, SISR for 3D Magnetics Resonance Images (MRI) is more challenging compared to 2D, mainly due to the increased number of neural network parameters, the larger memory requirement and the limited amount of available training data. 
    Current SISR methods for 3D volumetric images are based on Generative Adversarial Networks (GANs), especially Wasserstein GANs due to their training stability. Other common architectures in the 2D domain, e.g. transformer models, require large amounts of training data and are therefore not suitable for the limited 3D data. However, Wasserstein GANs can be problematic because they may not converge to a global optimum and thus produce blurry results. Here, we propose a new method for 3D SR based on the GAN framework. Specifically, we use instance noise to balance the GAN training. Furthermore, we use a relativistic GAN loss function and an updating feature extractor during the training process. We show that our method produces highly accurate results. We also show that we need very few training samples. In particular, we need less than 30 samples instead of thousands of training samples that are typically required in previous studies. Finally, we show improved out-of-sample results produced by our model.
\end{abstract}


\section{Introduction}
\label{sec:introduction}
% background
\IEEEPARstart{H}igh-resolution MR images are crucial for obtaining a detailed view of anatomical structures and are essential for downstream MRI analysis. However, the acquisition of high-resolution (HR) MRI is labor-intensive and susceptible to motion artifacts during the scanning process. Furthermore, reducing scan time often results in lower spatial resolution and the loss of fine details in the image structure. Therefore, the research focus in deep learning has shifted towards the Single Image Super Resolution (SISR) task, which aims to recover high-resolution MR images from lower-resolution images of the same subject.

As a result of the curse of dimensionality and limited availability of data, training an SISR model on 3D medical images is an ill-posed problem that is more challenging than training on common 2D images. Medical images, such as brain MRI, contain anatomical information in all three dimensions, making the task of SISR on 3D volumetric MRI the focus of this paper.

% Here it is talking about 2D and 3D advances
%In the 2D domain, numerous studies have been done to achieve high image fidelity in SISR tasks, \textit{e.g.}\cite{dong@srcnn,wang2018esrgan,srgan@ledig,vdsr}. These models are mostly trained using GAN architecture\cite{sGAN}, which is one of the most popular and powerful architectures in generative models. Recently, other models such as score-matching models, probability density models, and transformers have also produced SR results with advanced image quality in both metric scores and visual fidelity. Some of these models have been re-implemented in the 3D MRI domain, with the majority of them being GANs\cite{DCSRN,srmri,ArSSR,mri@gradient}, and they keep updating the {\emph{state-of-the-art}} performance.
In the 2D domain, several studies have been conducted to achieve high image fidelity in SISR tasks, \textit{e.g.} \cite{dong@srcnn,wang2018esrgan,srgan@ledig,vdsr}. Most of these models are trained using GAN architecture\cite{sGAN}, which is a popular and powerful architecture in generative models. Recently, other models such as score-matching models\cite{scorematching@song}, diffusion probability models\cite{DDPM}, and transformers\cite{vit} have also been used to produce SR results with advanced image quality in both metric scores and visual fidelity. In the 3D MRI domain, some of these models have been re-implemented, with the majority of them being GANs\cite{DCSRN,srmri,ArSSR,mri@gradient}, and they continue to improve the state-of-the-art performance.

% Here we start to mention problems in GANs, and describe our solutions
However, GAN training is known to be very unstable, sensitive to parameter changes, and requires careful design of the architecture, especially in the high dimension space and with deep architecture design. Wasserstein variations of GAN\cite{wgan@Arjovsky}\cite{wgangp@Ishaan} are the most effective and popular methods implemented in training SR models in the 3D MRI domain. Despite their effectiveness, we have observed that the results from the WGAN (or WGANGP) variations (e.g. DCSRN\cite{DCSRN}) are often prone to blur, which can be a consequence of their convergence to suboptimal points or oscillating around the global optimum\cite{dirac_gan,clcgan}.

In this paper, we propose to tackle this problem by introducing instance noise to the model and incorporating relativistic GAN loss\cite{ragan} into a three-player adversarial training\cite{tripleGAN} framework. Our proposed model exhibits stable training dynamics and produces high-quality results, demonstrating superior generalisation performance on previously unseen data compared to other models.

% -[ ] put into discussion, since OOD is not the main point of the paper
In the field of SR models on MRI, it is common practice to train on a fixed dataset without variation in modality or machine type. This approach can lead to limitations in model generalisability and domain-specific biases. However, in the case of brain MRI images, the shared content among images allows for effective inductive bias. This allows the model to learn informative features even with limited training data. 

To test this assumption, we trained a model on a single subject and successfully recovered the HR image. We further observed that training on dozens of samples did not sacrifice performance and our model exhibited superior generalisability on out-of-distribution (OOD) data, indicating minimal mode collapse in the GAN. To summarise, our contributions in this work are listed:
%Commonly the SR models are trained on a fixed dataset with no variation in types of modality or machine, leading to domain or modality-specific limitations and reduced model generalisability. However, since most brain MRI images share similar contents, we assume that training the model on limited data can still provide an effective inductive bias. 
% -[ ] rearrange the following sentence with"we have trained on few samples." 
% -[ ] another sentence only for OOD.
%To test the assumption, we have trained a model on just one subject and successfully recovered the HR image. We found that training on dozens of samples did not sacrifice performance and our model showed superior generalizability on out-of-distribution (OOD) data, indicating minimal mode collapse in the GAN. To summarise our contribution in this work:
\begin{itemize}
  \item we show empirically that our model exhibits better convergence towards the global optimum than other models on high-dimensional data(\textit{e.g.} 3D brain MRI).
  \item we show that the performance of our model remains robust even with a limited number of training samples.
  \item we show our model can generalize well to datasets of different modalities and resolutions.
  %validate on large public dataset
\end{itemize}

\begin{figure*}
    \centering
    \subcaptionbox{HR\label{fig:short-a}}{\includegraphics[height=6.18cm]{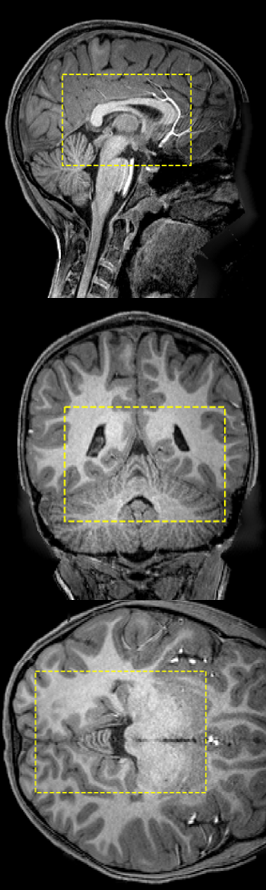}}%
    \subcaptionbox{ESRGAN \label{fig:short-b}}{\includegraphics[width=0.17\linewidth]{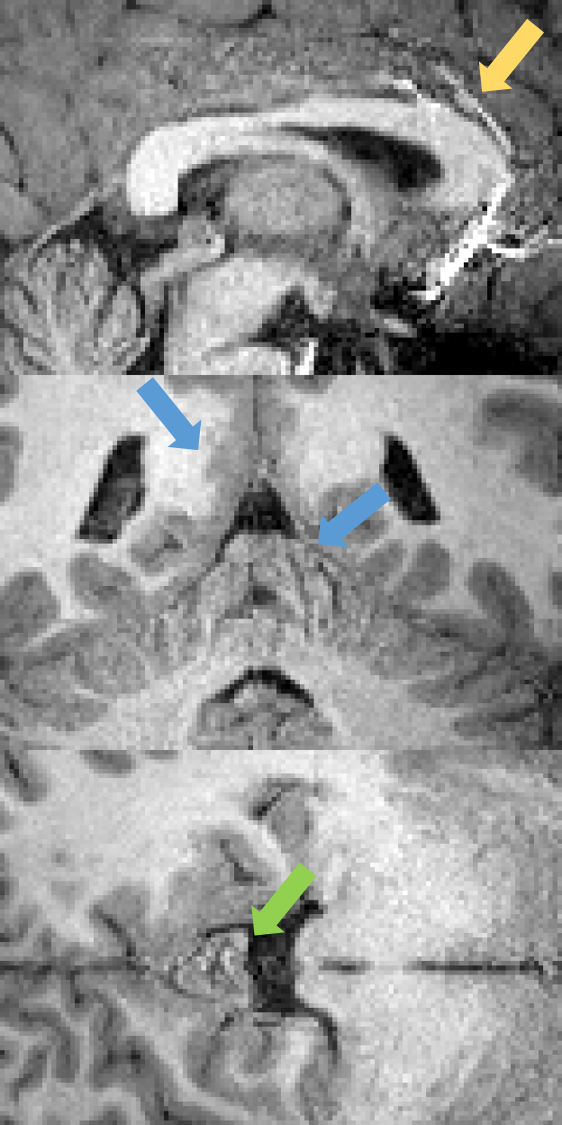}}%
    \subcaptionbox{ArSSR \label{fig:short-f}}{\includegraphics[width=0.17\linewidth]{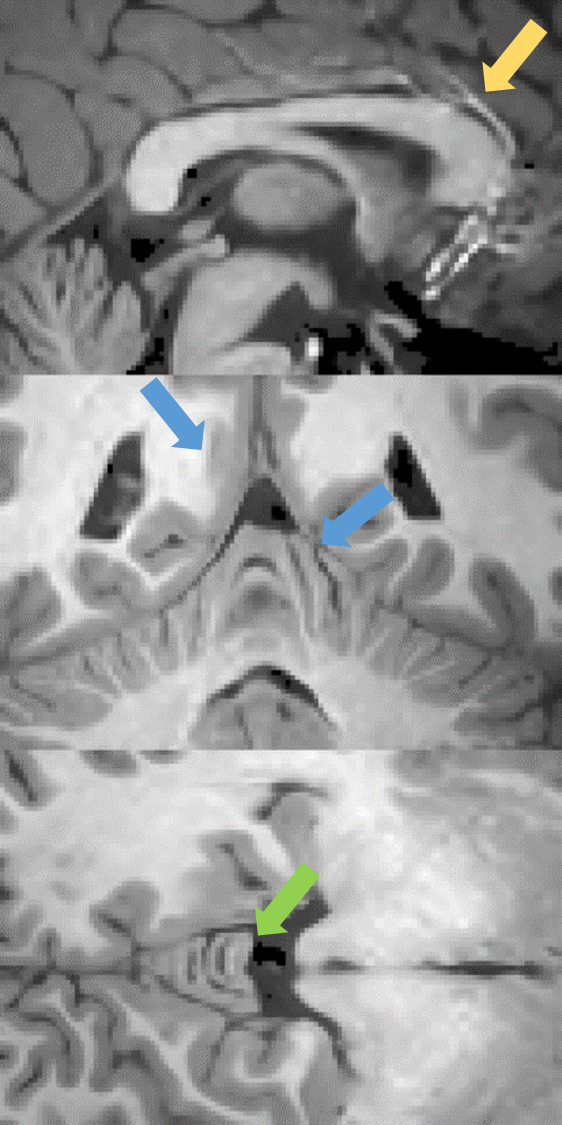}}%
    \subcaptionbox{DCSRN \label{fig:short-c}}{\includegraphics[width=0.17\linewidth]{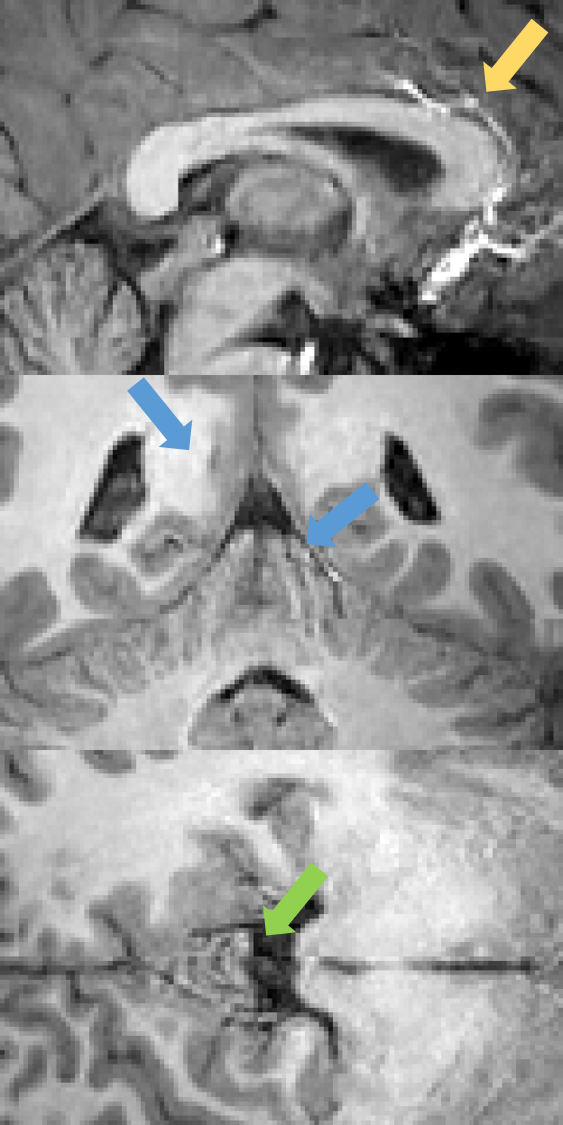}}%
    \subcaptionbox{\textbf{\textit{ours}} \label{fig:short-d}}
    {\includegraphics[width=0.17\linewidth]{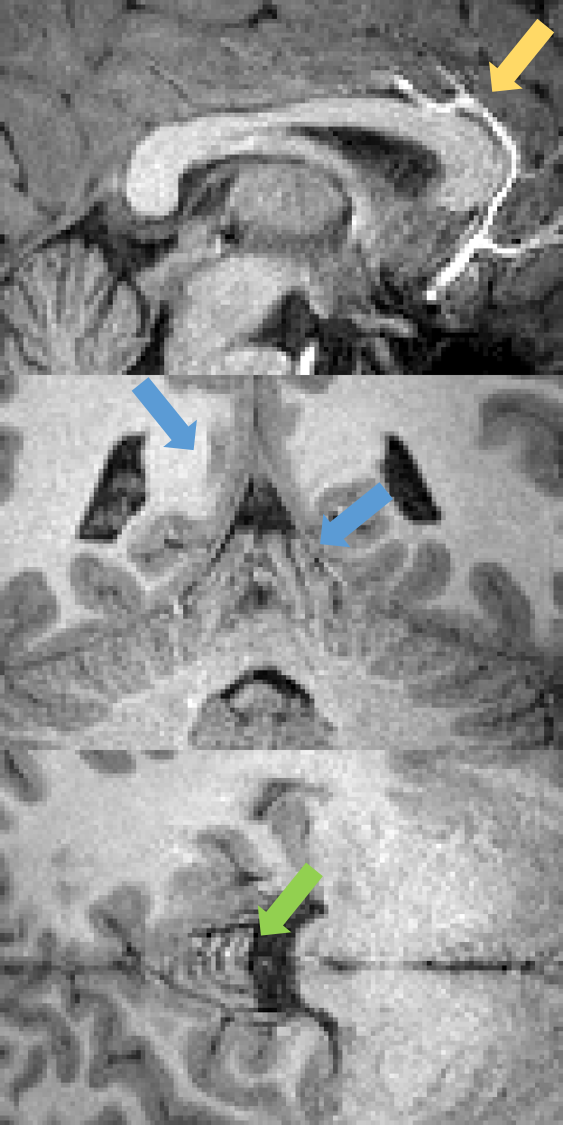}}%
    \subcaptionbox{GT \label{fig:short-e}}
    {\includegraphics[width=0.17\linewidth]{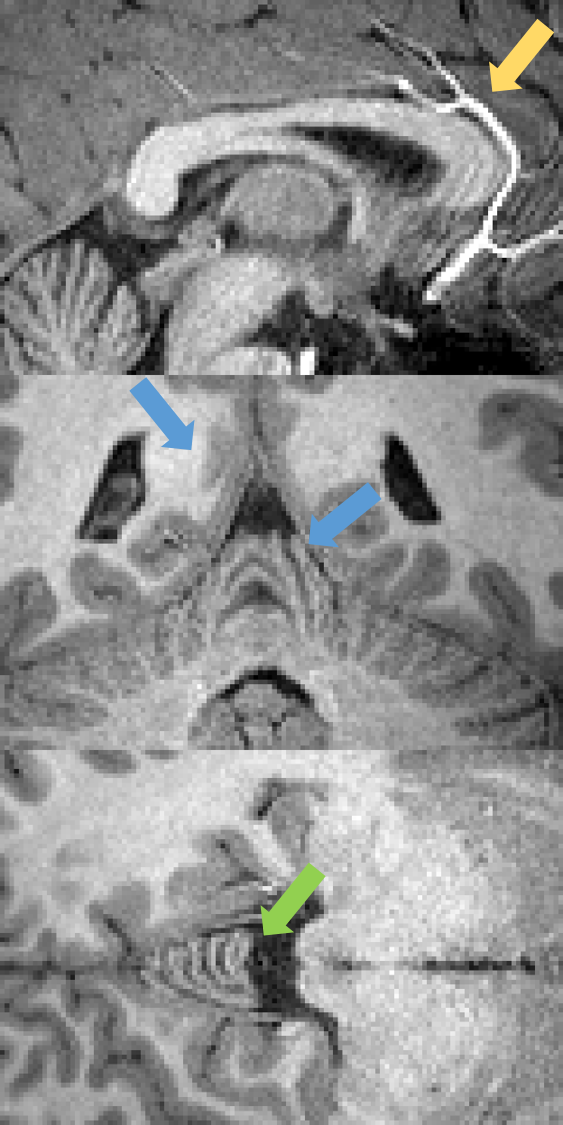}}%
    \caption{Qualitative comparison on an exemplary subject using different models with a $\times 2$ resolution upscale. \textbf{(a)} shows the High-resolution ground truth(HR) of whole brain MRI, where the \textit{yellow dashed box} highlights the location of the zoom-in views. The zoom-in views in subfigure\textbf{(b) - (f)} depictes the SR results generated by the following models: ESRGAN\cite{wang2018esrgan}, ArSSR\cite{ArSSR}, DCSRN\cite{DCSRN}, \textbf{ours}, and GT, in sub-figure\textbf{(b) - (f)} respectively. Our results show the best visual quality and the most detailed recovery of the GT.(Distinguishable differences are marked by \textit{orange, blue} and \textit{green} arrows in the sub-figures{\bf (b)-(f)}).}
    \label{fig:fig1}
  \end{figure*}

\section{Related Work}
\subsection{Super-Resolution in MR Images}
% 2D in common
Learning based SISR has been actively studied in both 2D and 3D images. In a SISR task, the neural network aims to learn a non-linear mapping from the low resolution image (LR) to its high resolution (HR) reference. This is particularly suited to the convolutional neural network (CNN) learning capability, as demonstrated in previous studies\cite{dong@srcnn,srgan@ledig,rcan,wang2018esrgan}. %cite

In the 2D domain, SRCNN\cite{dong@srcnn} was the first to propose training an end-to-end CNN architecture to capture the feature used for mapping LR to HR, demonstrating that the deep learning methods surpass the traditional SISR methods by a large margin. Subsequently, many architectures have been proposed to improve the performance of the models on SISR task. SRGAN\cite{srgan@ledig} was the first to employ a GAN framework, skipping connection\cite{resnet}, and perceptual loss\cite{feifeili} to achieve better perceptual quality of the produced SR images\cite{srgan@ledig}. The VDSR\cite{vdsr} model improved performance by using a deeper architecture that allowed larger model capacity. RCAN\cite{rcan} achieved superior performance by using channel attention and more layers to train an efficient SR network. ESRGAN\cite{wang2018esrgan} combined many efficient modules, including a pretrained feature extractor and relativistic GAN loss\cite{ragan}, to achieve SR results with high fidelity. Different from ESRGAN which used a frozen pretrained feature extractor, SPSR\cite{spsr} updated a third module to extract the gradient of the image as an additional constrain to generate higher fidelity results. Recently, transformers\cite{vit}\cite{transformer_sr} have been incorporated to perform \emph{state-of-the-art} quality in the SISR task. As a novel approach, diffusion models\cite{scorematching@song}\cite{sr3} are adopted to generate realistic and high quality data, with a stable training process and achieve better metric scores. These methods are all dominant in 2D SISR tasks, providing \textit{state-of-the-art} image quality while requiring large amounts of training data as well.

% 2D MRI
Many of these studies have inspired the development of models for SISR in medical imaging domain, some of which have achieved superior performance in SISR task. In the 2D MR image domain, \cite{MRSR_SEN} has used squeeze-excitation attention network to achieve remarkable super-resolution results in 2D MRI, while \cite{transformer_mr} adopted a transformer architecture to achieve superior image quality in a multiscale network. Despite the successful implementations in the 2D MR images, downstream analysis typically requires 3D volumes of MR images. Simply stacking 2D slices of SR results to create a 3D volume can lead to undesired artefacts along the cross-plane axes. Therefore, implementing a 3D model can directly outperform 2D models\cite{srmri}.

 This paper will focus only on 3D MR image training. Most of the novel architectures have not been adapted to 3D MRI data, due to the high dimensionality and lack of data. Thus, especially for 3D MRI domain, the GAN category remains the mainstream training method, given its efficiency and high performance. An initial work by Pham et al.\cite{srmri} discussed the implementation of CNN models on 3D MRI data for SISR tasks, demonstrating the advanced SR results produced by neural networks over traditional interpolation methods. In Chen et al.\cite{DCSRN}, the author combined WGAN training with densely connected residual blocks and reported prominent image quality. While in \cite{ArSSR}, the authors used implicit neural representation to realise SR by generating images at an arbitrary scale factor. In \cite{mri@gradient}, the authors proposed to use image gradient as a knowledge prior for the SR task, for a better matching to local image patterns.

Although many of these studies successfully produced SR results, they are mainly constrained to the same modality or imaging sequence as the training dataset. Also the quality of the reconstruction is sometimes limited.

\subsection{Stabilized Generative Adversarial Networks}
As was described above, GANs play an essential role in SISR. Typically, a GAN model consists of two models that compete adversarially against each other during the training process. In a standard GAN model, the generator network produces synthetic results that are as indistinguishable as possible from the real samples, while the discriminator network tries to identify whether the generated results are different from the real training samples or not. 

The objective function of the model is commonly referred to as the \textit{mini-max} game. During the training, the generator aims to maximise the probability of the discriminator classifying its generated samples as real, while the discriminator aims to minimise the probability of incorrectly classifying the generated samples as real. 

Due to the non-concave and non-convex optimisation property of GAN, it can be challenging to train and converge to the global optimum. One of the main causes of this is imbalanced training, where the discriminator converges too quickly, hindering the generator's ability to learn meaningful features. Much effort has been put into both stabilising training and improving convergence to the global optimum\cite{ragan}\cite{biggan}\cite{dirac_gan}\cite{clcgan}\cite{wgan@Arjovsky}\cite{wgangp@Ishaan}\cite{spectralnorm}. These modifications can be roughly categorised into four types: objective function, normalization layers, gradient regularization, and instance-wise regularization. 
One approach to stabilizing GAN training is weight clipping, coupled with the use of Wasserstein loss function, which was proposed in \cite{wgan@Arjovsky}. This approach was further improved in \cite{wgangp@Ishaan} by mixing both real and generated data as input for the discriminator, but still did not guarantee convergence to the optimum. In \cite{spectralnorm}, spectral normalization layers were added to improve stability, while in \cite{biggan}, orthogonal regularization was used to improve performance, stability, and scaling of the model. Mescheder et al.\cite{dirac_gan} conducted a theoretical investigation into the convergence properties of different GAN regularizations and found that weight-clipping GAN\cite{wgan@Arjovsky, wgangp@Ishaan} as well as the standard GAN\cite{sGAN} do not lead to convergence, but instead to local oscillations. This was further investigated by Xu et al.\cite{clcgan} from the perspective of control theory, where Xu et al. showed that convergence of Dirac GAN can be achieved by close-loop controlling. Amongst these prevailing regularizations, adding instance noise is shown to be the simplest and a computationally cheap way of stabilising the training, ensuring convergence, and providing high-quality image generation\cite{dirac_gan}\cite{instancenoise}.

\section{Our Approach}
Regarding the model design, we constructed a GAN comprising three networks - a discriminator, a generator, and a feature extractor - that were updated concurrently. To achieve optimal GAN performance, we utilised the relativistic GAN loss as our objective function and introduced Gaussian noise to enhance the model's stability. 

Specifically, the SISR model is trained using paired HR and LR inputs of the same subject, where the LR input is obtained by down-sampling the HR reference image by a fixed scale factor (\textit{e.g.}, $\times 0.5$ down-sampling in our training). This is formulated as $I_{LR} = f(I_{HR})$, where $f(\cdot)$ denotes the downgrading function. In our case, we use linear down-sampling, which results in information loss. Therefore, the reverse process (\textit{i.e.}, $g(\cdot) \approx f^{-1}(\cdot)$), known as the SR process, aims to recover high-resolution images from the LR images using a trained neural network in reference to the original HR images. The recovered SR image ($\hat{I}_{SR}$) is formulated as $\hat{I}_{SR} = g(I_{LR}) = f^{-1}(I_{LR})$. The reverse process $g(\cdot)$ is parameterized by a neural network, and the model continuously updates its parameters to learn the mapping between the distribution of the downgraded LR samples and the training HR samples.
\begin{figure}[H]
  \centering
  \includegraphics[width=0.8\linewidth]{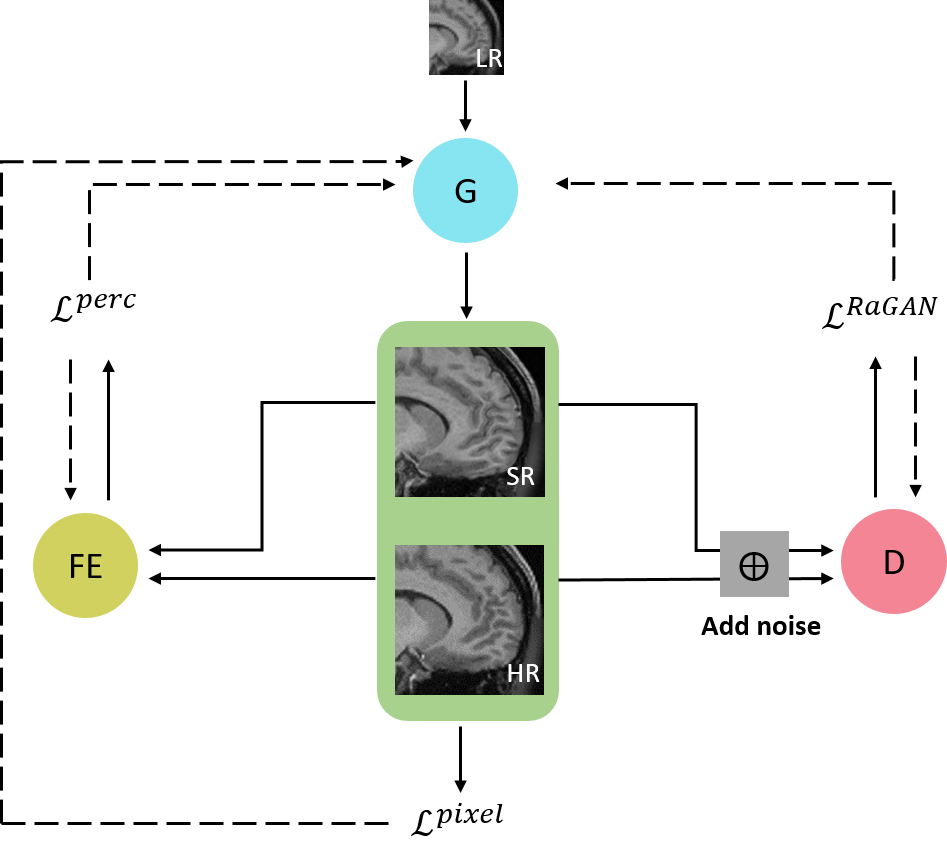}
  \caption{Semantic diagram of our network during training. The discriminator, generator, and feature extractor are denoted as D, G, and FE respectively (see pseudo code in Algorithm \ref{algo:1}). The \textit{solid} arrow denotes image data flow, while the \textit{dashed} arrow refers to the gradients which update the according networks; the \textit{gray} block denotes adding instance noise to the SR and HR images, the \textit{green} area denotes high-resolution image space. The low resolution(LR) MRI data is fed to the G network first, then SR MRI data is generated and is separately fed into D and FE networks. \textit{Note}: our feature extractor network updates concurrently with the generator.}
  \label{fig:diagram}
\end{figure}
% TODO:
%   1.simplify the subsections to not look like textbook, but in the form of:
%     -[x] we use XXX, XXX is defined as XXXXXXXXX, the discussion on XXXXXX  
\subsection{GAN Based SR Model}
We use the GAN architecture to train the SR model, which contains a generator, a discriminator, and a feature extractor. Typically, in a standard GAN, the input of the generator is a noise sample from a prior distribution, such as Gaussian distribution. However, in our context, the generator takes LR images as input and outputs super-resolution images, while the discriminator tries to distinguish between the generated images and the ground truth high-resolution images (see figure\ref{fig:diagram}). In this paper, we utilise the following notations: $D, G, FE$ represent the discriminator, generator, and feature extractor, respectively; $\psi, \theta, \phi$ refer to the weight matrixes of their respective parameters.

The generator network in our architecture includes densely connected residual blocks, which facilitate efficient training, as well as sub-pixel layers\cite{pixshuffle} for up-sampling image dimensions. Meanwhile, the discriminator contains convolutional blocks with strided convolutional layers and instance normalization. In order to generate images with higher perceptual quality, we also incorporated a feature extractor that constrains the Euclidean distance between the SR and HR images. Our overall architecture can be viewed as a multi-player differential game\cite{multiplayerGAN}, with the feature extractor acting as the third player, updated in an unsupervised manner instead of as a classifier. The gradient updates of the overall model are described in algorithm\ref{algo:1}.

% TODO:
% -[x] add more caption to introduce notations
% -[x] cross reference to the figure2  
\begin{algorithm*}
  \caption{Simultaneous gradient updating of our model. We use default values of learning rate $\gamma$=0.0001, coefficients of Adam optimiser $\beta_{1}$ = 0.9 and $\beta_{2}$ = 0.999, see section\ref{sec:implementation} and figure\ref{fig:diagram}.}\label{algo:1}
  \begin{algorithmic}[1]
  \Require Initialise parameters, $\psi, \theta, \phi$, of D, G, FE with Kaiming initialization. 
    \While{not converge}
    \State Sample $x\sim\mathbb{P}$, a batch from the real images.
    \State Sample annealed noise $\epsilon\sim\mathcal{N}(0,\sigma)$, with standard deviation $\sigma$ decreasing linearly from 1 to 0 per iteration.
    \State $\tilde{x}$ is from linearly down-sampling $x$.\Comment{Simulating LR image $\tilde{x}$}
    \State $y \gets G(\tilde{x})$ \Comment{Generate SR image $y$}
    \State $\mathcal{L}^{pixel}(x,y) \coloneqq \lvert x-y \rvert$ 
    \State $\mathcal{L}^{perc}(x,y) \coloneqq \lvert FE(x) - FE({y})\rvert$
    \State $\phi \gets$ Adam($\nabla_{\phi}[\mathcal{L}^{perc}(x,y)], \phi, \gamma, \beta_{1}, \beta{2}$) \Comment{Update feature extractor}
    \State $\theta \gets$ Adam($\nabla_{\theta}[\mathcal{L}^{perc}(x,y) +\alpha\mathcal{L}^{pixel}(x,y)+\beta\mathcal{L}^{RaGAN}_{G}(x+\epsilon, y+\epsilon)], \gamma, \beta_{1}, \beta_{2}$) \Comment{Update generator}
  
    \State $\psi \gets$ Adam($\nabla_{\psi}[\mathcal{L}^{RaGAN}_{D}(x+\epsilon, y+\epsilon)], \gamma, \beta_{1}, \beta_{2}$) \Comment{Update discriminator}
  \EndWhile
  \end{algorithmic}
  \end{algorithm*}

\subsection{Densely Connected Residual Blocks}
As the building block of the generator, we implemented densely connected residual blocks\cite{Huang_2017_CVPR}. While many studies have used deeper architectures to extract more informative latent features, deep architectures can be prone to instability during training. This can cause vanishing or exploding gradients. To address this concern, \cite{wang2018esrgan} and \cite{spsr} proposed to use densely connected residual blocks inside the generator models. This not only ensures the stability of the training but also improves the computational efficiency compared to generic ResNet\cite{resnet}. 

% -[x] refer to fig.2, and describe to which models are those loss function applied
% -[ ] discuss why use ResNet-10 not MSE nor VGG-19.(experimentally we found VGG-19 works worse than ResNet-10)

% -[x] introduce the L^{perc} at the beginning of the para. also refer to the figure2
% -[ ] add a sentence to differentiate VGG-perc and ours
% -[ ] move the discussion of using MSE to later context
\subsection{Perceptual Loss} 
To ensure the perceptual fidelity of the SR, we propose to use a perceptual loss function ($\mathcal{L}^{perc}$ in algorithm\ref{algo:1}) to constrain the quality of SR results in feature space. The perceptual loss was first introduced in \cite{srgan@ledig}, where a pretrained VGG-19 network\cite{VGG} was used to extract features. However, our implementation differs in that we do not use the pretrained VGG-19 network, but an actively updating ResNet10 network. More details are explained in the "Feature Matching" section below. 

 Our perceptual loss($\mathcal{L}^{perc}$ in algorithm\ref{algo:1}) is defined as a $l_{1}$ loss in feature space, see equation(\ref{eq:perc}). Here $I_{i,j,k}$ and $\hat{I}_{i,j,k}$ represent HR and SR intensity value at the $i,j,k$-th voxel in all three image dimensions($W,H,D$), $\mathcal{F}(\cdot)$ represent the feature network. 

\begin{equation}
  \mathcal{L}^{perc} = \frac{1}{W \times H \times D} \sum_{i,j,k=1}^{W,H,D}|\mathcal{F}(I_{i,j,k}) - \mathcal{F}(\hat{I}_{i,j,k})|
  \label{eq:perc}
\end{equation}

As shown in \cite{srgan@ledig}, perceptual loss functions generally outperform MSE loss functions, because they better preserve the detailed texture of the image. 

% @12:37  21 Feb 2023
\subsection{Relativistic Average GAN Loss}
As a key factor to characterise the GAN dynamics, we use relativistic GAN loss\cite{ragan} to make the training more efficient and stable. The standard GAN loss only discriminates between real and fake samples, but the relativistic average GAN loss function estimates how close the generated distribution is to the real distribution.  

Compared to the standard GAN loss, which converges slower and suffers from instability, the relativistic GAN loss allows the model to converge faster and show better convergence stability. The discriminator is encouraged to distinguish real data as being more realistic than the average of the generated data. The real samples are denoted as $x\sim\mathbb{P}$, with $\mathbb{P}$ the distribution of the real samples; the generated data are noted as $y\sim\mathbb{Q}$, with $\mathbb{Q}$ the distribution of the generated samples.
We propose to use the relativistic averaged discriminator($D_{Ra}$), where $C(\cdot)$ is the critic network(i.e. discriminator without the last sigmoid function, $\tau(\cdot)$).
% TODO:
% - [ ] The notations in eq.2,3,4 look tedious, try to make it concise, or even combine them into one
% -[ ] make an equation array to refer only one number of this block
\begin{equation}
  D_{Ra}(x,y) = \tau(C(x) - \mathbb{E}_{y \sim \mathbb{Q}}C(y))
\end{equation}
\begin{equation}
  D_{Ra}(y,x) = \tau(C(y) - \mathbb{E}_{x \sim \mathbb{P}}C(x))  
  \label{eq:ragan}
\end{equation}

The relativistic GAN loss for the discriminator($\mathcal{L}_{D}^{RaGAN}$) and the generator($\mathcal{L}_{G}^{RaGAN}$) is defined below. See also algorithm\ref{algo:1}.
\begin{equation}
  \begin{split}
    % RaGAN loss for D    
    \mathcal{L}_{D}^{RaGAN}(x,y) = & -\mathbb{E}_{x\sim \mathbb{P}}
    [\log(D_{Ra}(x,y))]\\
    &- \mathbb{E}_{y \sim \mathbb{Q}}
    [\log(1 - D_{Ra}(x,y)]
  \end{split}
  \label{eq:ragan_d}
\end{equation}

\begin{equation}
  \begin{split}  
  % RaGAN loss for G
    \mathcal{L}_{G}^{RaGAN}(x,y) = & -\mathbb{E}_{x\sim \mathbb{P}}
    [\log(1- D_{Ra}(y,x))] \\
    & - \mathbb{E}_{y \sim \mathbb{Q}}
    [\log(D_{Ra}(y,x)]
  \end{split}
\end{equation}

\subsection{Instance Noise}
As noted before, GANs suffer from convergence problems. Therefore we add linearly annealed Gaussian noise to the input of the discriminator to avoid early phase divergence, as proposed in \cite{instancenoise}. 

Adding instance noise to both the generated SR samples and their reference samples before feeding them to the discriminator stabilises the training process. The main concept is to make the real and the generated distributions overlap by adding Gaussian noise. This has been shown to be an effective and computationally cheap way of balancing the GAN dynamics and ensuring convergence\cite{instancenoise}. The expression for the discriminator can be rewritten as follows: 

\begin{equation}\label{eq:inst}
  D(I + \epsilon, \hat{I} + \epsilon)
\end{equation}
where $\epsilon \sim \mathcal{N}(0, \sigma)$ is the Gaussian noise with linearly decreasing variance from 1 to 0, with each update of the network gradients.

\subsection{Feature Matching}
In order to enhance the quality of generated samples, we employed feature matching, which involves updating the feature extractor network alongside the generator. This technique has been shown to improve the perceptual quality of images in previous studies\cite{feifeili, srgan@ledig}. Typically, a VGG-like network\cite{VGG} is used as the feature extractor due to its pretrained ability to capture meaningful feature representations\cite{wang2018esrgan}. However, since we did not have access to any pre-trained networks for our application, we developed our own feature matching network.
%To improve the quality of generated samples, we applied feature matching by simultaneously updating the feature extractor network together with the generator. As was proposed in previous studies\cite{feifeili,srgan@ledig}, feature matching has been shown to be an effective way in improving the perceptual quality of images. Most of the studies use a VGG-like network\cite{VGG} as feature extractor, because a well-pretrained VGG on a natural image dataset captures meaningful feature representations\cite{wang2018esrgan}. However, to the best of our knowledge, pretrained networks are not available at present, we developed our feature-matching network. 

We implemented a three-player game\cite{multiplayerGAN} where a feature extractor network is randomly initialised and updated alongside the generator. The gradient indirectly updates the feature extractor, with the objective function being the perceptual loss. The generator and feature extractor work in pairs to improve the perceptual quality of the image in an end-to-end manner. To serve as the feature extractor module, we utilized the lower layers of ResNet10 before the multi-perceptron layers, and updated its weights from random initialization during the GAN training.

While standard perceptual loss setups use the first 54 layers of a pre-trained VGG19 network for feature extraction\cite{wang2018esrgan}, we found that ResNet10, a shallower architecture, performed better in our experiments when updated without pretraining. Additionally, when using the VGG-like feature extractor, we noticed that the generated images had undesirable checkerboard artefacts, which were not present when ResNet10 was used. Overall, the best quality of generated samples was achieved when concurrently updating ResNet10 with the generator to minimise the perceptual loss.

\subsection{Objective Function}
In order to generate images with both high perceptual quality and balanced training performance, we include $\mathcal{L}_{1}$ loss in the image domain, perceptual loss, and GAN loss as the overall objective function for the generator network. The parameters of the generator are constrained to learn both feature-wise ($\mathcal{L}^{perc}$) and image-wise ($\mathcal{L}^{pixel}$) representations of the ground truth, which the objective function seeks to minimise. The overall objective function of the generator is defined as:

\begin{equation}\label{eq:objfn}
  \mathcal{L}_{G} = \mathcal{L}^{perc} + \alpha \mathcal{L}^{pixel} + \beta \mathcal{L}_{G}^{RaGAN}
\end{equation}
where $\alpha = 0.01$ and $\beta = 0.005$ are empirical parameters for weighting image space $\mathcal{L}_{1}$ loss and Relativistic averaged GAN loss. 

The discriminator and the feature extracting networks are updated based on their respective loss, $\mathcal{L}_{D}^{RaGAN}$ and $\mathcal{L}^{perc}$, as defined in equation\eqref{eq:ragan_d} and equation\eqref{eq:perc}. 
% @13:56  21 Feb 2023
\section{Experimental setup}
% TODO:
% -[x] Change the names of dataset to indicate their usage: dataset-contrast, dataset-resolution, etc.
% -[x] add " " for all dataset names
\subsection{Datasets}
To test our approach, we used several datasets from different scanners, different imaging modalities or parts of body. Three of those datasets are part of the Human Connectome Project\cite{david2013hcp}, the rest is from \cite{knee}. We name the datasets by the most distinguishable attribute of the each dataset, as described below.

\textbf{Dataset "Insample":}
This dataset is downloaded from the \emph{Lifespan Pilot Project}, as the part of the Human Connectome Project (HCP). It consists of T1-weighted (T1w) MRI images from 27 healthy subjects aged 8 to 75 (15 females) acquired in a Siemens 3T scanner. In our model we split them as 20 subjects for training and 7 for test. The resolution of the ground truth images is $0.8 mm$ isotropic, with matrix size of $208\times 300\times 320$. 

\textbf{Dataset "Contrast":}
This dataset is also from the \emph{Lifespan Pilot Project} described above, but now we use the T2-weighted(T2w) images of the same subjects. 

\textbf{Dataset "Resolution":}
This test dataset consists of 1113 T1w brain MRI images, from the WU-Minn Young Adult study of the HCP, acquired in the 3T scanner at the resolution of $0.7 mm$ isotropic. 

\textbf{Dataset "Knee":}
This dataset contains 20 knee MRI proton density weight images downloaded from \cite{knee}.

% TODO:
% -[ ] Add the inference process, e.g. how to assemble image patchs
\subsection{Implementation Details} \label{sec:implementation}

In the implementation of the network, we trained a generator network, a critic network, and a feature extractor network. The generator network consists of 3 Residual-in-Residual Dense Blocks\cite{wang2018esrgan}, which are embedded densely connected residual units, without any batch norm layer. The critic network is the same as a discriminator without the final non-linear activation layer. The feature extractor uses convolutional layers of ResNet10 before linear layers. All three networks are initialised by Kaiming initialization\cite{kaiming_init} and are optimised by Adam optimizer\cite{adam}, using coefficients of $\beta_{1} = 0.9$, $\beta_{2} = 0.999$, and learning rate of $\gamma = 10^{-4}$. The variance of the instance noise decreases linearly in each iteration, from $\sigma = 1$ to $0$. The training simultaneously updates, on PyTorch framework\cite{PyTorch}, a feature extractor, a generator and a discriminator network for 60000 iterations, on a NVIDIA's Ampere 100 GPU. 

\subsection{Evaluation Metrics}
Several metrics have been proposed for measuring the accuracy of super-resolution results, for example, Structure-Similarity-Index-Measurement (SSIM), Peak-Signal-Noise-Ratio (PSNR), and Normalized-Root-Mean-Square-Error (NRMSE). However, these metrics sometimes do not reflect image fidelity correctly\cite{MA,LPIPS}. Since SISR is considered an ill-posed problem, these measurements only reveal the statistical similarity between the target and the reference, which are imprecisely high when the target is blurry so that the mean and variance of the target fit the ground truth better\cite{MA}. Therefore, here we also used Learned-Perceptual-Image-Patch-Similarity (LPIPS) metrics which better capture perceptual similarity\cite{LPIPS}. Here we use the 2D slice-wise implementation provided by \cite{LPIPS}. 

Perceptual metrics such as LPIPS have been pretrained on large image datasets and are known to extract meaningful features that closely resemble human visual quality \cite{MA, LPIPS}. However, unlike traditional metrics such as PSNR and SSIM, LPIPS may not always correlate with them. %As illustrated in Figure \ref{fig:blur} and Table \ref{table:psnr_dataset_a}, our approach obtained lower PSNR and SSIM values compared to others, yet exhibited higher quality in terms of details and sharper edges as indicated by the higher LPIPS score.
% -[ ] put result of ArSSR in dataset_C in the table II
\begin{figure*}
  \begin{center}
  \subcaptionbox{Ours (1 subject)\label{fig:1subj}}{\includegraphics[width=0.3\linewidth]{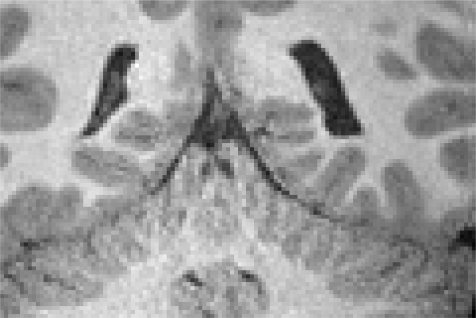}}%
  \subcaptionbox{Ours (20 subject)\label{fig:20subj}}{\includegraphics[width=0.3\linewidth]{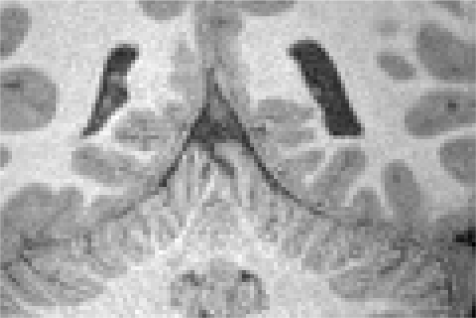}}%
  \subcaptionbox{GT\label{fig:GT}}{\includegraphics[width=0.3\linewidth]{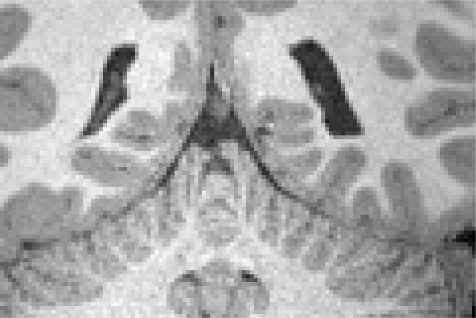}}%  
  \end{center}
  \caption{Effect of training dataset size on the quality of super-resolution (SR) images. From \textit{left} to \textit{right} are the zoom-in views of the SR images generated by the models trained on: 1 subject, 20 subjects, and ground truth. \emph{Note}: The brain structure is mostly preserved in all subfigures except for the contrast change in subfigure\textbf{(a)}. Our proposed model achieves better perceptual quality in the cerebellum when trained on 20 subjects in subfigure\textbf{(b)}, indicating only slight improvement in SR quality with an increase in training samples}% 
  \label{fig:numbersubj}%
\end{figure*}%
\begin{figure*}
\begin{center}
  \subcaptionbox{ArSSR\label{fig:res_arssr}}{\includegraphics[width=0.19\linewidth]
  {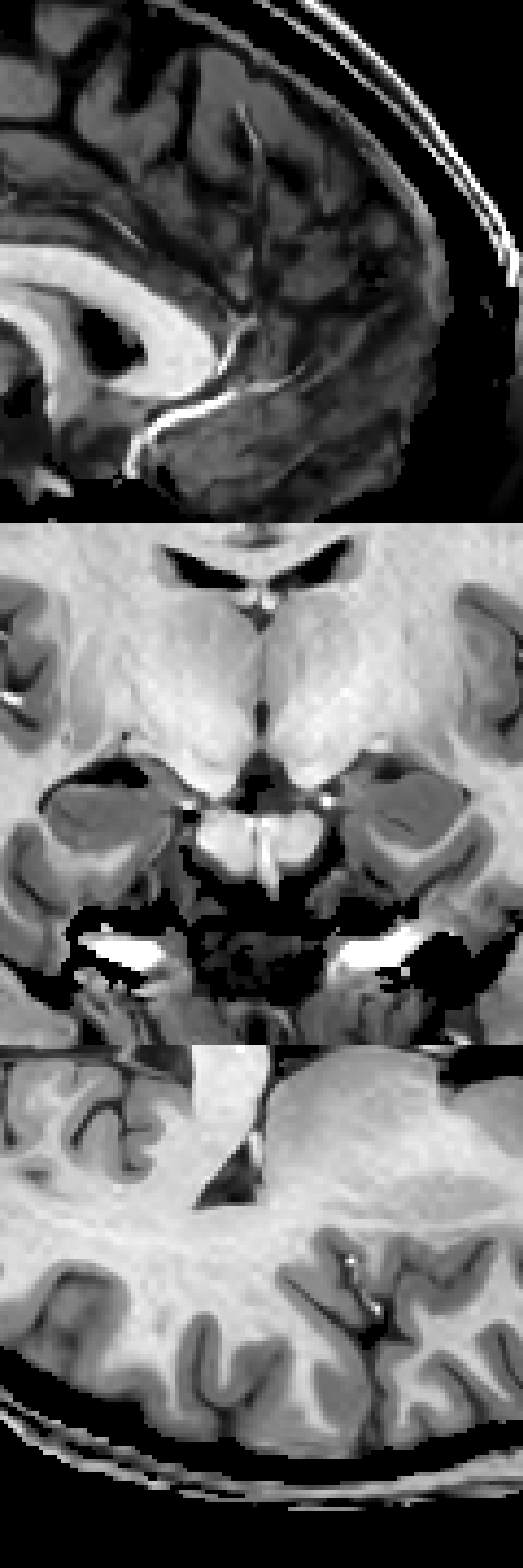}}%
  \subcaptionbox{DCSRN\label{fig:res_dcsrn}}{\includegraphics[width=0.19\linewidth]
  {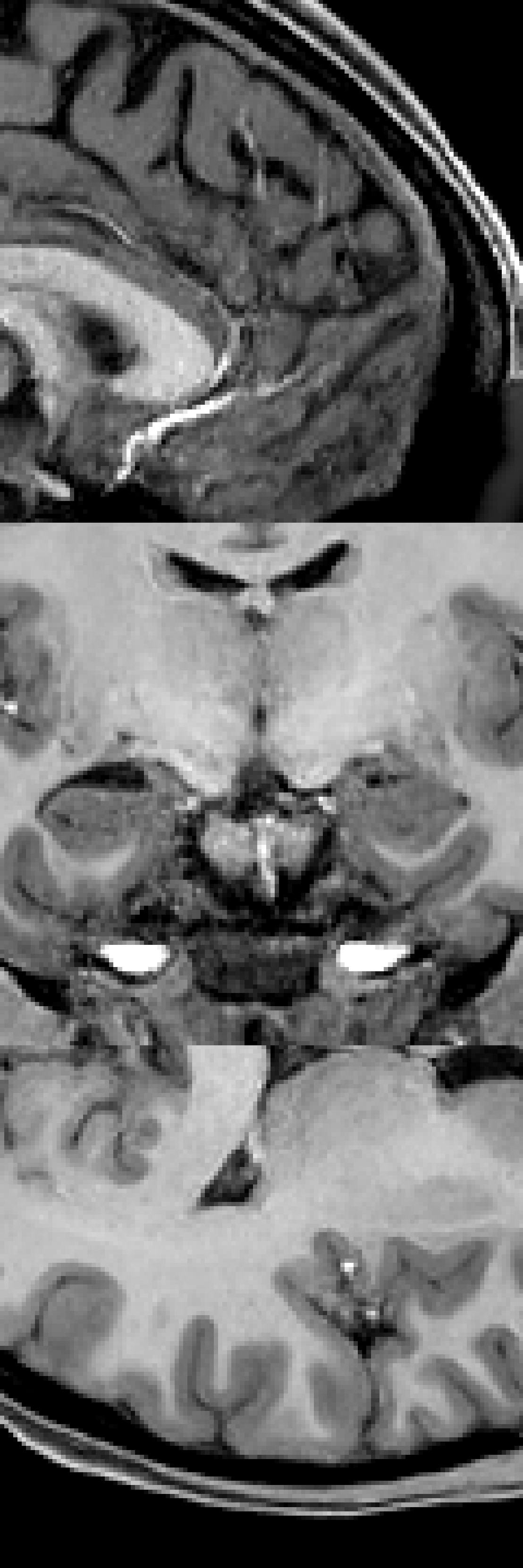}}%
  \subcaptionbox{ESRGAN\label{fig:res_esrgan}}{\includegraphics[width=0.19\linewidth]
  {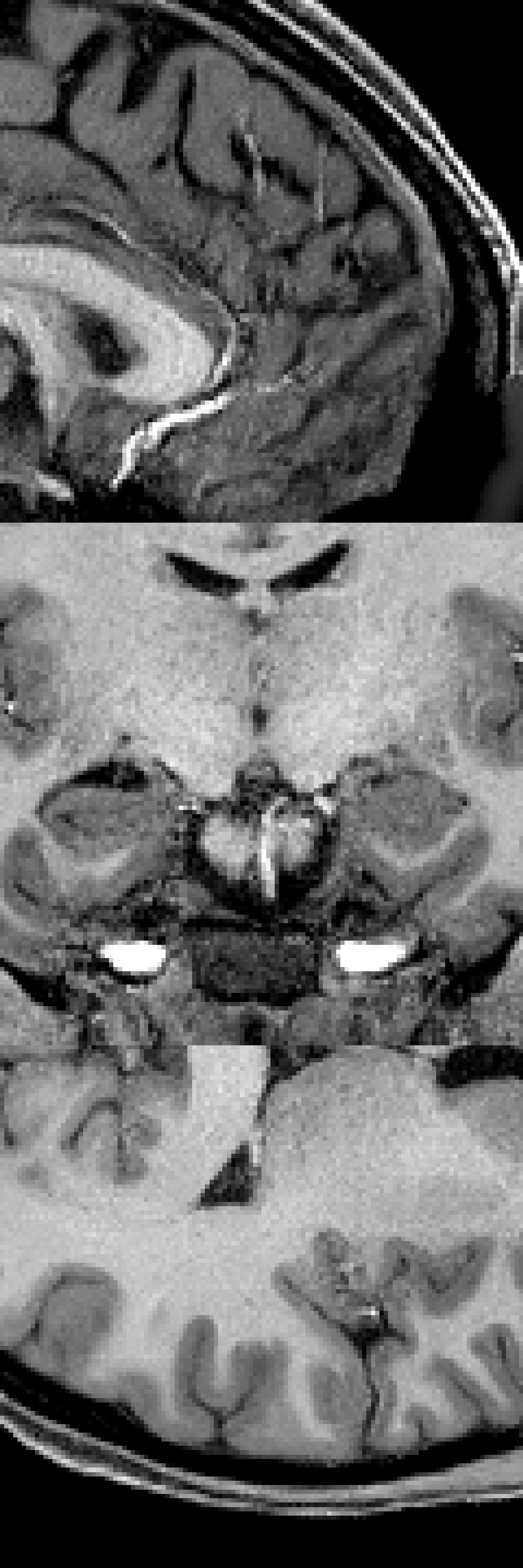}}%
  \subcaptionbox{\textbf{Ours}\label{fig:res_ours}}{\includegraphics[width=0.19\linewidth]
  {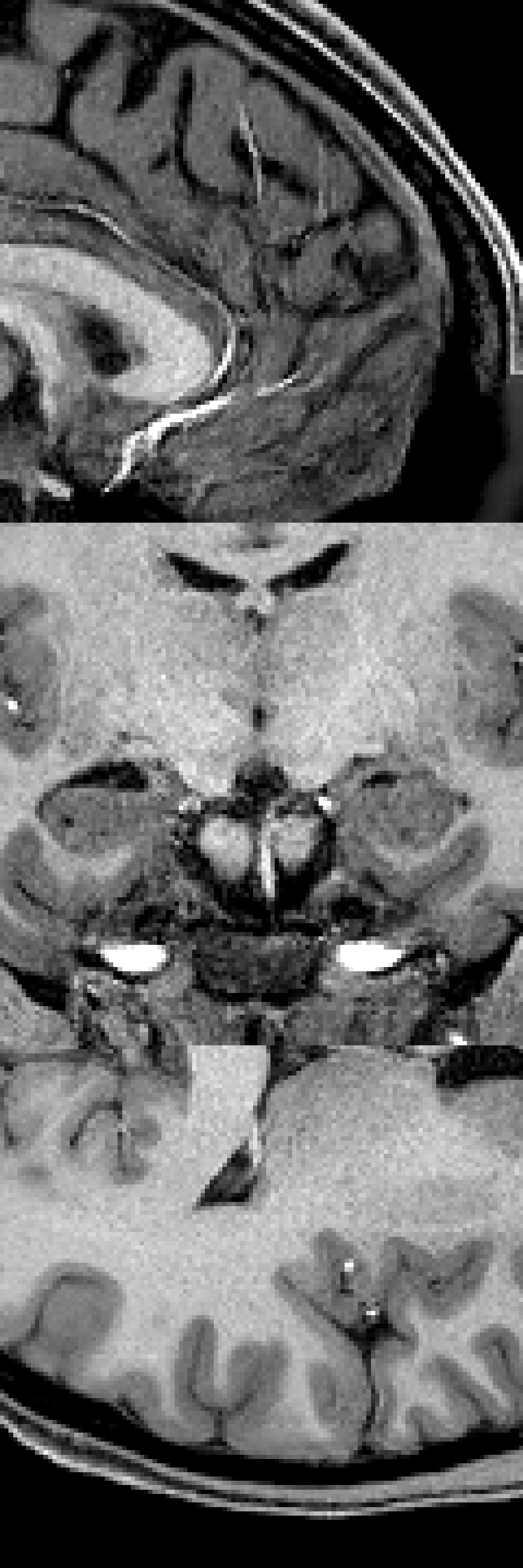}}%
  \subcaptionbox{GT\label{fig:res_gt}}{\includegraphics[width=0.19\linewidth]{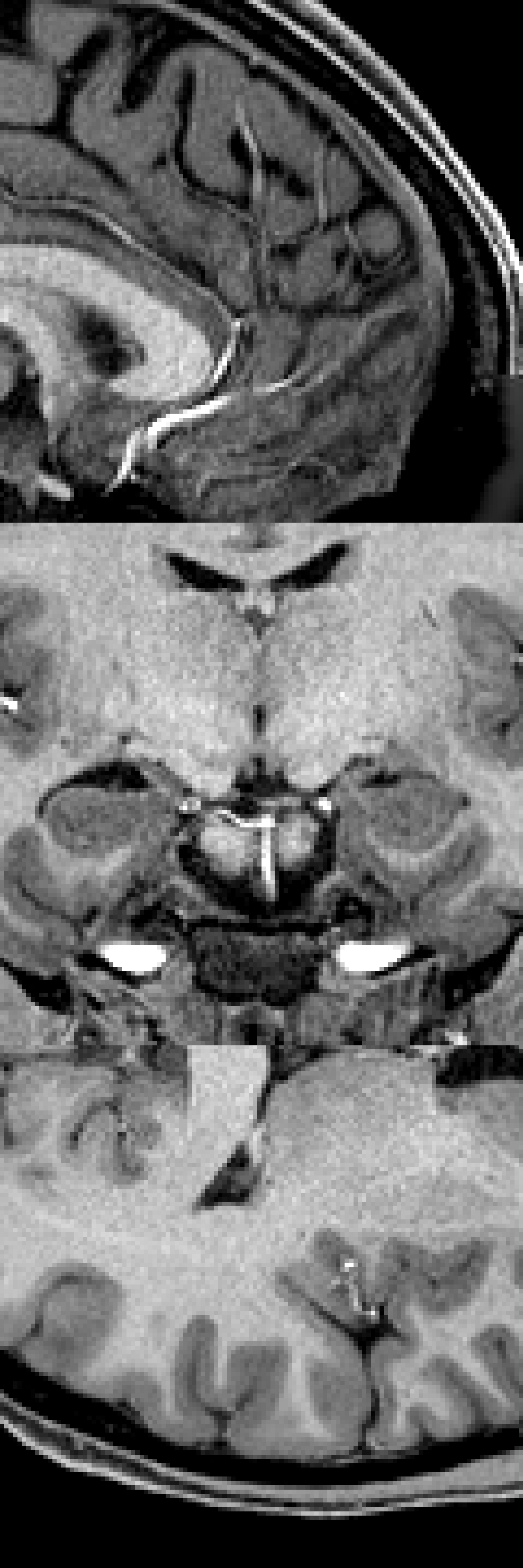}}%  
\end{center}
\caption{Qualitative comparison on dataset "Resolution" among zoom-in SR results with different perceptual quality. (a)-(e) shows visual comparison in a zoom-in view of SR results, in \emph{Dataset "Resolution"}: ArSSR, DCSRN, ESRGAN, \textbf{Ours}, and GT. \textit{Note}: Our result shows the best quality visually and a lowest LPIPS score, despite of having low PSNR and SSIM score.
}
\label{fig:blur}
\end{figure*}

% HOW TO:
% 1. write what was done in this experiment, 
% 2. write what and how other baseline models are configured
% 3. describe results referred to figures and tables

\section{Results}
\subsection{Experiment 1: Insample Super-Resolution}
%Constraining gradient of the network explicitly can result in blurry image quality, while these methods are popular to stabilize GAN training.
The first experiment is to test in-sample super-resolution using \emph{Dataset "Insample"}. Due to the memory constraints, we split the images into overlapping patches of size $64\times 64\times 64$, which are defined by sliding a window in step size 16 along each dimension with zero-padding at the end. For each subject, 2160 patches with overlapping patches were obtained as training data. To simulate lower-resolution images, LR patches were obtained by linearly down-sampling HR patches to half of their original resolution, namely to a resolution of $1.6 mm$ with a matrix size of $32\times 32\times 32$. The paired patches of LR and HR images were fed into the model for training. 

% -[x] description of baseline models  
For comparison, we tested our approach against three competing approaches, namely ESRGAN\cite{wang2018esrgan}, ArSSR\cite{ArSSR}, and DCSRN\cite{DCSRN}. ESRGAN was only available in 2D, and we reconfigure it in 3D and retrained it on a 3D dataset. Specifically, we pretrained the first 54 layers of a 3D VGG-19 network on 3D MRI images. Both the ArSSR and DCSRN models are available in 3D, but we re-trained the model on \emph{Dataset "Insample"}. 

The results are shown in figure\ref{fig:fig1}. Note that our model recovered the ground truth best, especially with regard to small blood vessels (subfigure\ref{fig:short-d}); our model also recovered very detailed cerebellum structures (second and third row of figure\ref{fig:fig1}). The quantitative results are shown in table\ref{table:psnr_dataset_a}. Note that our method outperformed other models in NRMSE and LPIPS. 

\begin{table}[h]
    \begin{center}
    \begin{tabular}{@{}l c c c c c@{}} 
     \toprule
     Model            & PSNR $\uparrow$ & SSIM $\uparrow$ & NRMSE $\downarrow$ &  LPIPS(2D)$\downarrow$\\ %[0.5ex] 
     \midrule
     Tri-linear       & 33.038 & 0.876 & 0.023 & 0.084 \\
     ESRGAN-3D        & 37.022 & 0.933 & 0.013 &0.044\\ 
     DCSRN            & \textbf{37.635} & \textbf{0.954} & 0.013 &0.052 \\
     ArSSR	      & 28.038 & 0.280 & 0.048 & 0.291\\
     \textbf{\textit{ours}} & 36.922 & 0.943 & \textbf{0.013} &\textbf{0.040}\\ %[1ex] 
     \bottomrule
    \end{tabular}
    
  \end{center}
    \caption{Quantitative comparison of performance among the \textit{state-of-the-art} models on \textit{dataset "Insample"}, where $\uparrow$ indicates the higher the value the better the image quality, \textit{vice versa} for $\downarrow$. \textit{Note:} the sharpness and LPIPS value better represent image visual quality in accordance to the figure\ref{fig:blur}.}
    \label{table:psnr_dataset_a}
\end{table}%
\subsection{Experiment 2: Small Training Samples}
In this experiment, we trained our model on \emph{Dataset "Insample"} using only one image for training. Here we compared the performance of models trained on 1 and 20 subjects, the qualitative performance of the 1 subject-trained model is on par with the 20 subject-trained one as is shown in figure\ref{fig:numbersubj}.

% -[ ] the following should go to discussion
Owing to the common properties of brain MRI, which contains less informative background and relatively less heterogeneous sample contents, convolutional neural networks can efficiently learn meaningful feature representations of brain MRI patches for SR tasks\cite{MRSR_SEN}. Given that we assume it is possible for a 3D convolutional neural network to learn strong enough inductive bias towards SISR, using a limited amout of training samples.

Our model utilizes the relativistic GAN and feature matching techniques to significantly improve efficiency. This intrinsic property of brain MRI data means that fewer training samples are required compared to common 2D vision tasks, making it possible to reduce the amount of training data needed for the brain MRI model. Empirically, we demonstrate that it is possible to address the data deficiency issue for MRI SR by training our model on just one subject. Interestingly, we observe that the generator's performance is not sensitive to the number of training samples when using the feature extractor network, such as ResNet10. Furthermore, during our experiments, we found that a deeper architecture used for feature matching does not outperform a shallower one. For instance, ResNet50 performs worse than using ResNet10 as a feature extractor. 

We noticed the performance of the generator is not sensitive to the numbers of the training samples, when using the feature extractor network(\textit{e.g.} ResNet10). 
%TODO
% -[ ] put below paragraph to feature extractor part
Besides, the deeper architecture used for feature matching does not perform better than a shallow one during our experiment (\textit{e.g.} ResNet50 performs worse than using ResNet10 as feature extractor).

\begin{figure*}[htp]
  \begin{center} 
  \subcaptionbox{Overview GT\label{fig:short-knee-overall}}{\includegraphics[width=0.16\linewidth]{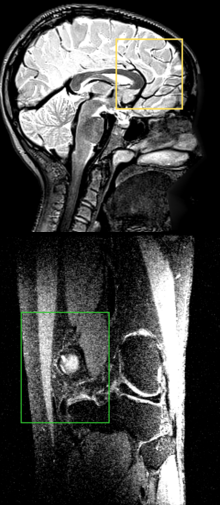}}%  
  \subcaptionbox{ESRGAN\label{fig:short-knee-esrgan}}{\includegraphics[width=0.16\linewidth]{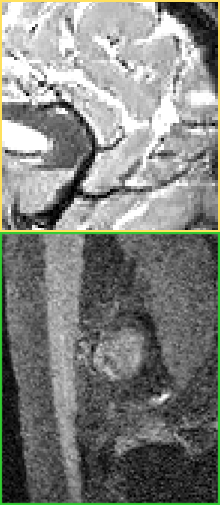}}%  
  \subcaptionbox{ArSSR\label{fig:short-knee-esrgan}}{\includegraphics[width=0.16\linewidth]{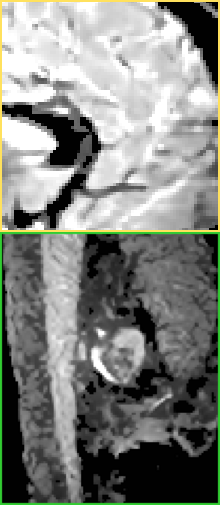}}%
  \subcaptionbox{DCSRN\label{fig:short-knee-DCSRN}}{\includegraphics[width=0.16\linewidth]{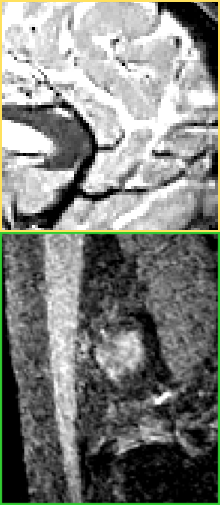}}%
  \subcaptionbox{\textbf{Ours}\label{fig:short-knee-ours}}{\includegraphics[width=0.16\linewidth]{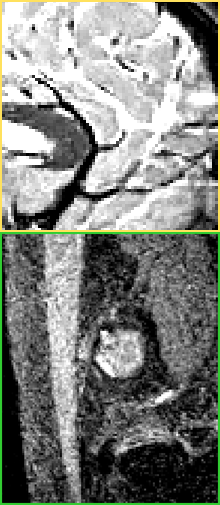}}%
  \subcaptionbox{GT\label{fig:short-knee-gt}}{\includegraphics[width=0.16\linewidth]{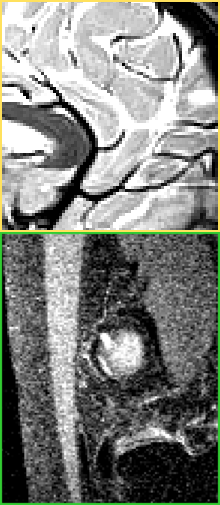}}% 
  
  \end{center}
  \caption{Qualitative comparison on out-of-distribution datasets "Contrast" and "Knee" for different models. Each column shows: (a) overall view of GT image and (b)-(f) zoom-in SR results of ESRGAN\cite{wang2018esrgan}, ArSSR\cite{ArSSR}, DCSRN\cite{DCSRN}, \textbf{Ours}, and ground truth; The \textit{top} row shows the SR results of models tested on brain MRI images with T2w contrast from dataset "Contrast". Our model produces the most accurate vessel structures.; the \textit{bottom} row shows the SR results of the models tested on knee MRI images with PDw contrasts from dataset "Knee". Our model outperformed the other models in reconstructing detailed knee joint and surrounding tissue structures.}
  \label{fig:ood}
\end{figure*}
\subsection{Experiment 3: Zero-Shot Inference on OOD Data}

In our out-of-distribution experiment, we trained our model using the Dataset "Insample" and evaluated its performance on three different datasets: "Contrast", "Resolution", and "Knee". The ground truth of these OOD datasets exhibits significant variations in terms of contrast, resolution, and body part, as are shown in figure\ref{fig:blur} and \ref{fig:ood}.

For comparison purposes, we also tested ESRGAN, DCSRN, and ArSSR, trained on the dataset "Insample", on these OOD tasks. Our model outperformed the other models in this experiment on all three datasets, by best producing image details that most closely approximated the GT. 

Our model performed better in recovering the details of vessels in the sagittal plane of the SR image, as shown in the top row of figure\ref{fig:blur} on the "Resolution" dataset. In contrast, other models, such as ArSSR and DCSRN, produced over-smoothed images, while ESRGAN produced images with noisy details. Our model also exhibited superior image quality on the "Contrast" and "Knee" datasets, where it closely approximated the blood vessels and the round-shaped bone structure, as demonstrated in subfigure\ref{fig:short-knee-ours}.

%Shown in the results, the details of the vessels in the sagittal plane of the SR image were better recovered by our model, on dataset "Resolution" as in the \textit{top row} of figure\ref{fig:blur}. Whereas, the rest of the models either produce over-smoothed, \textit{e.g.} in ArSSR and DCSRN, or noisy image details, \textit{e.g.} ESRGAN. The superior image quality is also shown when evaluating our model on datasets "Contrast" and "Knee", where blood vessels and the round-shaped bone structure were closely approximated by our model, as demonstrated in figure\ref{fig:short-knee-ours}.

Although DCSRN obtained the highest PNSR and SSIM, the LPIPS score of our model was well aligned with its image quality and details, as demonstrated in figure\ref{fig:blur} and table~\ref{table:psnr_dataset_a} and~\ref{table:psnr_dataset_c}.

\begin{table}[h]
  \begin{center}
  \begin{tabular}{l c c c c c} 
   \toprule
   Model & PSNR $\uparrow$ & SSIM $\uparrow$ & NRMSE $\downarrow$ & LPIPS$\downarrow$\\ %[0.5ex] 
   \midrule
   ESRGAN-3D(20 subj) & 37.181 & 0.957 & 0.013 &0.039\\ 
   DCSRN(1 subj) & 37.420 & 0.959 & 0.013 &0.088 \\
   DCSRN(20 subj) & \textbf{37.564} & \textbf{0.962} & \textbf{0.013} &0.051\\
   ours(1 subj) & 36.355 & 0.953 & 0.014 &0.048\\
   ours(20 subj) & 36.922 & 0.953 & \textbf{0.013} &\textbf{0.038}\\ %[1ex] 
   \bottomrule
  \end{tabular}
  
\end{center}
  \caption{Quantitative comparison of robustness among the \textit{state-of-the-art} models on {\bf \emph{Dataset "Resolution"}}.}
  \label{table:psnr_dataset_c}
\end{table}

Our experiments showed that our approach can be generalized to different datasets with little decrease in performance. Notably, even when exposed to vastly different contrast or anatomy, our model did not collapse into specific training data and generalized well on out-of-distribution datasets.
    
%\begin{table}[h]
%\begin{adjustbox}{width=\columnwidth, center}
%  
%  \begin{tabular}{@{}l c c c c c@{}} 
%	      \toprule
%       Data & PSNR $\uparrow$ & SSIM $\uparrow$ & NRMSE %$\downarrow$ &LPIPS(2D)\\ %[0.5ex] 
%       \midrule
%       {\bf \emph{Dataset "Contrast"}} & 33.522 & 0.923 & 0.016  &%0.056\\ %[1ex] 
%       {\bf \emph{Dataset "Knee"}} & 38.455 & 0.873 & 0.014  &0.%114\\ %[1ex] 
%       \bottomrule
%      \end{tabular}
%      \captionsetup[table]{skip=10pt}
%    \end{adjustbox}
%      \caption{Quantitative illustration of OOD generalizability %of our model on T2w brain MRI and knee MRI.}
%      \label{table:ood}
%    \end{table}
    \section{Discussion}
    In this paper, an effective method is proposed for stabilizing GAN training, which outperforms \emph{state-of-the-art} models in SR tasks on 3D brain MRI data. Our approach involves adding Gaussian noise to discriminator inputs, using a three-player GAN model, and applying the relativistic GAN loss as the objective function. Our experimental results demonstrate that this method can ensure effective training convergence and achieve better generalizability. Several key findings are presented in this paper. First, the proposed model generates SR results with superior perceptual quality compared to existing methods, even when the amount of training data is small. Secondly, the model achieves accurate SR generation and convergence even when trained on a single subject. Finally, the model exhibits strong generalizability and is less prone to overfitting, as evidenced by its successful performance on previously unseen datasets (\textit{i.e.} \textit{Dataset "Contrast"}, \textit{Dataset "Resolution"}, and \textit{Dataset "Knee"}).

%    % TODO:
    % -[ ] not writing too much limitation on the model itself, when it's not certain to be a limit. 
    %Despite achieving competitive results, our model has room for improvement in addressing the following open questions. Notably, our model utilized a shallow network for feature matching, which surprisingly outperformed a deeper one. This counter-intuitive behavior highlights the need for further research to better understand the role of feature extractor design in three-player games.

    While the LPIPS score is a reliable metric for assessing the perceptual quality of images, it is primarily designed for 2D image evaluation and may overlook information in the cross-plane of volumetric MRI. Therefore, future research should focus on developing specialized methods for evaluating the perceptual quality of 3D MRI images.

    %\subsection*{Acknowledgment}
    %The authors would like to acknowledge the use of ChatGPT for polishing the style and the grammar of the English writing.
    %{\small
    \bibliographystyle{ieee_fullname}
    \bibliography{SR_TMI_DeepL}
    %}
    
    \end{document}